# A Derivation of the Colburn Analogy

Trinh, Khanh Tuoc

K.T.Trinh@massey.ac.nz

## Abstract

This paper presents a derivation of the empirical Colburn analogy and discusses its implications.

Key words: heat transfer, Colburn, j factor, turbulence, power law, log-law, wall layer

## 1 Introduction

The study of heat and mass transfer in turbulent flows has been heavily influenced by the formulation of an analogy with the better known law for momentum transfer by Reynolds (1874).

$$St = \frac{f}{2} \tag{1}$$

where

$$St = \frac{h}{\rho C_p V} \text{ is called the Stanton number,} \tag{2}$$

$$f = \frac{2\tau_w}{\rho V^2} \text{ the friction factor,} \tag{3}$$

$C_p$ the thermal capacity and

$V$ the average fluid velocity

$h$ the heat transfer coefficient

This analogy works well for the region outside the relatively thin layer where from the wall diffusion is important but the exact connection between the wall and the outer region has been elusive (Trinh, 2010b). Thus a number of empirical correlations were proposed early on to provide useful tools for engineering design. The most famous is perhaps the Colburn analogy (1933).

$$St = \frac{f}{2} Pr^{-2/3} \qquad (4)$$

$$j = St\, Pr^{2/3} = \frac{f}{2} \qquad (5)$$

where

$$Pr = \frac{C_p \mu}{k} \text{ is the Prandtl number} \qquad (6)$$

$k$     the fluid thermal conductivity

$\mu$     the fluid viscosity, and

Though good semi-theoretical correlations were later derived (Metzner and Friend, 1958a, Trinh, 1969) the Colburn analogy is still often quoted and its j factor often used.

This paper presents a theoretical derivation of the Colburn analogy and highlights its implications.

## 2    Theory

The velocity and temperature profiles can be expressed in terms of a log-law and a power law. The power law form is

$$\frac{U}{U_\infty} = \left(\frac{y}{\delta_{m,t}}\right)^p \qquad (7)$$

$$\frac{\Theta - \Theta_w}{\Theta_\infty - \Theta_w} = \left(\frac{y}{\delta_{h,t}}\right)^{p'} \qquad (8)$$

where $U$ and $\Theta$ are the local time averaged velocity and temperature and the suffices $w$, $\infty$ refer to values at the wall and at an infinite normal distance in fact at the edge of the boundary layer.

The logarithmic form is

$$U^+ = 2.5\ln y^+ + B \qquad (9)$$

$$\Theta^+ = 2.5\ln y^+ + B' \qquad (10)$$

where $U^+ = U/u_*$, $y^+ = yu_*\rho/\mu = yu_*/\nu$ and $\Theta^+ = \rho C_p (\Theta - \Theta_w) u_*/q_w$ have been normalised with the friction velocity $u_* = \sqrt{\tau_w/\rho}$, $\rho$ the fluid density, q the rate of heat transfer flux and y is the normal distance from the wall.

There is still a debate as to which representation should be used (Afzal, 2001, 2005; Zagarola & Smits, 1998). Power law profiles have a major weakness: they result in a zero velocity and temperature gradient at the wall and cannot be used to predict the rate of momentum and heat transport at the wall. We side step that problem by assuming that equations **Error! Reference source not found.**) and **Error! Reference source not found.**8) do not apply straight to the wall and that a thin layer exists that is dominated by a diffusion process. This restriction is compatible with modern observations of the turbulent process (Kline et al., 1967) where heat, mass and momentum are transferred by conduction to a thin unsteady state wall layer during a so-called inrush phase and then by convection as the fluid in the wall layer is ejected in a violent burst. Since the ejected fluid contain both heat, mass and momentum the convection process is the same for all and an analogy is justified but only in the region outside the wall layer (Trinh, 2010b, Trinh et al., 2010).

Thus to complete the picture, we force equations (7) and (8) through a known point in the flow field. Three points are traditionally taken: the edge of the wall layer $\delta_h^+, \theta_h^+$, the edge of the thermal buffer layer $\delta_{b,h}^+, \theta_{b,h}^+$ which was shown to be the time-averaged value of $\delta_h^+$ (Trinh, 2009) and $\delta_{k,h}^+, \theta_{k,h}^+$ the edge of the diffusive sub-layers postulated by Prandtl. The form of the final correlation depends on the choice of the reference point: $\delta_{b,h}^+, \theta_{b,h}^+$ yields the Karman (1939) Martinelli (1947) analogies; $\delta_{k,h}^+, \theta_{k,h}^+$ the Prandtl-Taylor (1910) and Meztner-Friend (1958b) analogies.

In this analysis, we pass equation (7) through the Kolmogorov point which is the intersection of the log law with the equation

$$U^+ = y^+ \qquad (11)$$

describing pure steady state diffusion. This is a fictitious point (Trinh, 2009) used simply for convenience e.g. by Levich (1962), Trinh (1969), Wilson and Thomas (1985) and many others. The coordinates of this Kolmogoroff point are $y_k^+ = U_k^+ = 11.8$. Substituting into equation (7) and rearranging we obtain

$$\delta_{m,t}^+ = 11.8 \left( \frac{U_\infty^+}{11.8} \right)^{1/p} \tag{12}$$

Now the intersection between equation (10) and the diffusive equation at the wall

$$\Theta^+ = y^+ \Pr \tag{13}$$

is $(11.8 \Pr^{-b}, 11.8 \Pr^{1-b})$ (Trinh, 2010b). Substituting into equation (10) gives

$$\delta_{h,t}^+ = 11.8 \Pr^{-b} \left( \frac{\Theta_\infty^+}{11.8 \Pr^{1-b}} \right)^{1/p'} \tag{14}$$

Then the ratio between the turbulent thermal and momentum boundary layers $\delta_{h,t}^+, \delta_{m,t}^+$ becomes

$$\sigma_t = \frac{\delta_{h,t}^+}{\delta_{m,t}^+} = \frac{\Theta_\infty^{+1/p'}}{U_\infty^{+1/p}} \Pr^{-b - \frac{1-b}{p'}} 11.8^{\frac{1}{p} - \frac{1}{p'}} \tag{15}$$

We now apply Reynolds' analogy by stating

$$p = p' \tag{16}$$

Then

$$\sigma_t = \left( \frac{\Theta_\infty^+}{U_\infty^+} \right)^{1/p} \Pr^{\frac{b - bp - 1}{p}} \tag{17}$$

By definition, for a boundary layer

$$\frac{\Theta_\infty^+}{U_\infty^+} = \frac{f/2}{St} \tag{18}$$

and

$$\sigma_t = \left( \frac{f/2}{St} \right)^{1/p} \Pr^{\frac{b - bp - 1}{p}} \tag{19}$$

The Stanton number is obtained from the integral energy equation (Knudsen & Katz, 1958)p.420)

$$St = \frac{\partial}{\partial x} \int_0^{\delta_{h,t}} \frac{U}{U_\infty} \left[ 1 - \left( \frac{\Theta - \Theta_w}{\Theta_\infty - \Theta_w} \right) \right] dy \tag{20}$$

Substituting for (7) and (8)

$$St = \frac{\partial}{\partial x}\int_0^{\delta_{h,t}}\left(\frac{y}{\delta_{m,t}}\right)^p\left[1-\left(\frac{y}{\delta_{h,t}}\right)^{p'}\right]dy \tag{21}$$

$$St = \frac{\partial}{\partial x}\left[\frac{p\delta_{m,t}}{(1+p)(1+p+p')}\sigma_t^{p+1}\right] = \left[\frac{p\sigma_t^{p+1}}{(2p+1)(p+1)}\right]\frac{\partial \delta_{m,t}}{\partial x} \tag{22}$$

Most experimental data show that the ratio $(f/2)/St$ is independent of $x$ (Reynolds claimed that it is unity, 1874) and equation (19) indicates then that we can take $\sigma_t$ to be independent of $x$. Then

$$St = \left(\frac{\delta_{m,t}}{x}\right)\left[\frac{p\sigma_t^{p+1}}{(2p+1)(p+1)}\right] \tag{23}$$

The boundary layer $\delta_{m,t}$ can be estimated from the integral momentum equation by standard techniques. The friction factor can be expressed as

$$f = \frac{\alpha}{\text{Re}_g^\beta} \tag{24}$$

giving (Skelland and Sampson, 1973, Trinh, 2010a)

$$p = \frac{\beta}{2-\beta} \tag{25}$$

and

$$\frac{\delta_{m,t}}{x} = \left[\frac{\alpha(\beta+1)(\beta+2)}{\beta(2-\beta)}\right]^{\frac{1}{\beta+1}}\left[\frac{\nu}{U_\infty x}\right]^{\frac{\beta}{\beta+1}} \tag{26}$$

$$\frac{\delta_{m,t}}{x} = \left[\frac{\alpha(1+3p)(1+2p)}{2p}\right]^{\frac{1+p}{1+3p}}\left[\frac{\nu}{U_\infty x}\right]^{\frac{2p}{1+3p}} \tag{27}$$

Combining (19), (23) and (27), and rearranging gives

$$St = \frac{f}{2}\Pr^{(b-bp-1)\frac{p+1}{p+2}} \tag{28}$$

Putting $b = 1/3$ for high Schmidt numbers, (Trinh, 2010b) and $p = 1/7$ (Blasius, 1913) gives

$$St = \frac{f}{2}\Pr^{-.63} \tag{29}$$

It is interesting to note that if we substitute equation (29) into (19) we get $\sigma_t \approx 1$ whereas the ratio $\sigma_l$ of the thermal to momentum boundary layer thicknesses in laminar flow is roughly $\sigma_l \approx \mathrm{Pr}^{-1/3}$.

## 3    Conclusion

The Colburn analogy can be derived from first principles if we assume that the index of the power law correlations for the velocity and temperature distributions is the same.